\documentclass[12pt]{article}

\usepackage[utf8]{inputenc}
\usepackage[T1]{fontenc}

\usepackage[top=1in,bottom=1in, left=1in, right=1in]{geometry}
\usepackage{amsmath, amsthm, amssymb, amsfonts, dsfont}

\usepackage{tabularx}
\usepackage{bbm}
\usepackage{mathtools}
\mathtoolsset{showonlyrefs}

\usepackage{subcaption}
\usepackage{graphics}
\usepackage{graphicx}
\graphicspath{{Figures/}}

\usepackage{numprint}
\npthousandsep{,}\npthousandthpartsep{}\npdecimalsign{.}

\usepackage{enumitem}
\usepackage[english]{babel}

\usepackage[sort,compress]{natbib}

\usepackage[x11names]{xcolor}
\usepackage{relsize}
\usepackage{breqn}
\usepackage{comment}

\usepackage{caption,url}
\usepackage[colorlinks=true,linkcolor=cyan,citecolor=blue]{hyperref}
\usepackage{booktabs}
\usepackage{stackengine}
\usepackage{authblk}
\usepackage{verbatim}

\usepackage{bm}
\usepackage{bbold}
\usepackage{blkarray}
\usepackage{multirow}
\usepackage{mwe}

\usepackage[ruled,vlined]{algorithm2e}

\usepackage[textsize=tiny]{todonotes}
\DeclareMathOperator*{\argmax}{argmax}
\renewcommand{\vec}[1]{\boldsymbol{#1}}

\usepackage{authblk}

\title{Hierarchical excitatory processes for modelling event-time data in the presence of exogenous stimuli}

\author[1]{Francesco~Sanna~Passino}
\author[1]{Nicholas~A.~Heard}
\author[2]{Jeffrey~W.~Brown}
\author[3]{William~N.~Frost}
\author[4]{Vince~P.~Lyzinski}

\affil[1]{Department of Mathematics, Imperial College London, London (United Kingdom)}
\affil[2]{Department of Biobehavioral Health, College of Health and Human Development, The~Pennsylvania~State~University, University~Park, Pennsylvania~(United~States)}
\affil[3]{Stanson~Toshok~Center~for~Brain~Function~and~Repair, Rosalind~Franklin University~of~Medicine~and~Science, North Chicago, Illinois (United States)}
\affil[4]{Department of Mathematics, University of Maryland, College Park, Maryland (United States)}

\date{}

\begin{document}

\maketitle

\begin{abstract}
We introduce the Hierarchical Excitatory Process (HEP), a flexible point process model for event-time data observed under repeated external stimuli. The proposed framework models the conditional intensity of a point process as a superposition of excitation effects induced by external stimuli, characterised by kernels with parameters dynamically evolving over time. This hierarchical construction enables modulation of excitation strength across repeated stimuli, providing an interpretable structure.
We establish likelihood-based inference for the proposed model and embed HEP within a model-based clustering framework to identify latent groups sharing similar response dynamics. Simulation studies demonstrate the model’s ability to recover evolving latent  patterns, and an application to spike train recordings from the sea slug \textit{Aplysia} pedal ganglion illustrates how HEPs are able to characterise stimulus-driven excitability of neurons across repeated stimulation under different experimental conditions.
\end{abstract}

\section{Introduction}
\label{sec:intro}

Event-time data arise in a wide range of scientific domains, including neuroscience \citep{Brown02}, epidemiology \citep{Meyer11}, finance \citep{Bowsher07}, and the social sciences \citep{Mohler11}. In many applications, events occur in the presence of 
exogenous interventions that cause temporary deviations from a baseline. 
For example, in neuroscience, neuronal spike trains could exhibit clear increases in firing intensity after an external stimulation, followed by gradual decay towards the baseline. However, both the magnitude and persistence of these responses often decrease following successive stimuli, reflecting phenomena such as habituation \citep[see, for example,][]{Thompson66}. 
Beyond neuroscience, similar dynamics can arise in several other domains. In public health, for example, advisories, such as a warning about an emerging infectious disease, can generate a sharp increase in events related to testing, clinic visits, or hospital admissions. Event intensities could return towards baseline as the public attention decreases, with the response varying across regions or population groups \citep[for example,][demonstrate an example of this phenomenon with the 2009 H1N1v influenza epidemic in England]{BrooksPollock11}.

A central modelling challenge in such phenomena is to characterise how such exogenous stimuli affect event intensities over time, while accounting for non-stationarity, adaptation, and heterogeneity across experimental units. 
To address this, in this work we propose \emph{Hierarchical Excitatory Processes (HEPs)}, a class of models for event-time data in the presence of exogenous stimuli. The model comprises a baseline activity, combined with stimulus-driven excitations induced via an explicit kernel representation, whose parameters are themselves allowed to vary dynamically between stimuli according to a point process model. This hierarchical construction provides a simple and interpretable way to model non-stationarity induced by exogenous changepoints, adaptation over repeated external interventions, and heterogeneity across experimental units.
The proposed framework yields interpretable parameters under specific kernel choices, describing excitability, persistence, and recovery dynamics after repeated exogenous stimuli. The HEP modelling framework also naturally supports model-based clustering through shared hierarchical structure. Estimation and inference are developed within a likelihood-based framework, and we demonstrate the approach using neuronal spiking data recorded under repeated external stimulation, with the objective of identifying groups of neurons that exhibit comparable patterns of activity over time. 

Classical point process models mainly offer two mechanisms for modelling clustering of events: piecewise deterministic time-varying intensities, such as inhomogeneous Poisson processes, and history-dependent models, such as self-exciting processes \citep[for example, Hawkes processes; see][]{Hawkes71}. These models are not directly designed to accommodate systematic changes based on exogenous stimuli, nor to capture changes to the excitation dynamics across repeated interventions. 
HEP combines ideas from these approaches by utilising a inhomogeneous Poisson process, comprising excitation effects accumulated from the external stimuli, with time-varying parameters for the coefficients of the excitation kernels. 
As a motivating example, we apply our proposed methodology to spike train recordings from the sea slug \textit{Aplysia} pedal ganglion. 
Within the context of spike train data, the use of point process models is well established in the literature, with generalized linear models commonly used to estimate relationships between event times and a set of covariates \citep[see, for example,][]{Eden2022}. 
HEP provides a flexible framework that may offer practitioners useful tools to \emph{(i)} quantify the effect of past exogenous interventions, \emph{(ii)} estimate the impact of additional future stimuli, \emph{(iii)} identify groups of neurons exhibiting similar response patterns, and \emph{(iv)} assess whether individual neurons display evidence of habituation. 

Within the context of temporal point processes, there have been some earlier attempts at incorporating external interventions within various modelling frameworks. For example, \cite{Zhang20} proposes a methodology to identify the set of exogenous events from a set of unlabelled events. The distinction between the role of endogenous and exogenous events in point processes has been primarily studied within the context of modelling social media. For example, \cite{Rizoiu17} adds exogenous effects within Hawkes-type models for popularity of online content. Similarly, \cite{Farajtabar17} includes exogenous control terms within multivariate Hawkes processes for the spread of fake news.  \cite{De18} separates between endogenous and exogenous effects for opinion dynamics. Within different applied contexts, \cite{Dong23} considers exogenous influences of city landmarks within a model for the spread of COVID-19, \cite{Zhu22} proposes a spatio-temporal attention point processes model for traffic congestion incorporating exogenous factors, focusing on police response to traffic
incidents, and \cite{Ruggeri25} considers a self-exciting mode for terrorist events with exogenous factors, interpreted as major economic or political events.

Beyond temporal point processes, the proposed procedure has links with the literature on causal mediation, which deals with the analysis of direct and indirect effects of external interventions on a response variable, especially in time-varying settings \citep[see, for example,][]{VanderWeele16, Aalen20, Hizli23}. While HEP does not explicitly identify causal effects in the counterfactual sense, its hierarchical structure assumes that repeated interventions modulate the strength of an intermediate excitation process that in turn governs the event intensity. Furthermore, the excitation kernels used in HEP can be interpreted as responses analogous to those arising in compartmental dynamical systems \citep[see, for example,][]{Li04,Salway08}. In this sense, each stimulus induces a latent response, similar to a pharmacokinetic concentration curve, which then decays over time, with decay rates dependent on the previous interventions. Unlike compartmental models, however, HEP treats this response at the level of the conditional intensity function for event-time data and allows adaptation of the response across repeated stimuli.

The rest of the paper is structured as follows: Section~\ref{sec:data} introduces the \textit{Aplysia} neuronal spiking data, illustrating the ``stylized facts'' motivating the modelling choices proposed in this work. Next, Section~\ref{sec:hep} presents hierarchical excitatory processes (HEPs), the main contribution of this work, along with associated likelihood-based inferential procedures and model-based clustering approaches. The performance of the model is examined via extensive simulation studies in Section~\ref{sec:simulations}, followed by the application to the \textit{Aplysia} data in Section~\ref{sec:applications}.

\section{Motivating example: \textit{Aplysia} neuronal spiking data} 
\label{sec:data}

As a motivating example for the proposed methodology, we consider optical recordings of neuronal activity in the dorsal pedal ganglia of the sea slug \textit{Aplysia californica}, a well-established model organism for investigating cellular mechanisms of learning and memory \citep[see, for example,][]{Kandel01, Bedecarrats20}. The dorsal pedal ganglia are part of the central nervous system, which in \textit{Aplysia} contains approximately 10,000 neurons. Recordings were performed on the isolated central nervous system, allowing the activity of multiple neurons to be captured before, during, and after the application of electrical stimuli. 
The chosen stimuli evoke stereotyped rhythmic escape locomotor programs, which are reflected as characteristic fluctuations and bursts in neuronal activity. The optical recordings provide signals from which the timing of spike events corresponding to neuronal firing can be inferred. By observing these continuous signals over time, the recordings capture the dynamic response of the neuronal population to repeated stimulation, providing insight into processes such as habituation, where neurons gradually adjust their activity in response to predictable stimuli. Full details about the data collection methodology are given in \cite{Hill20}.

\begin{figure}[t]
\centering
\begin{subfigure}{0.475\textwidth}
\centering
\caption{December 2022}
\includegraphics[width=\textwidth]{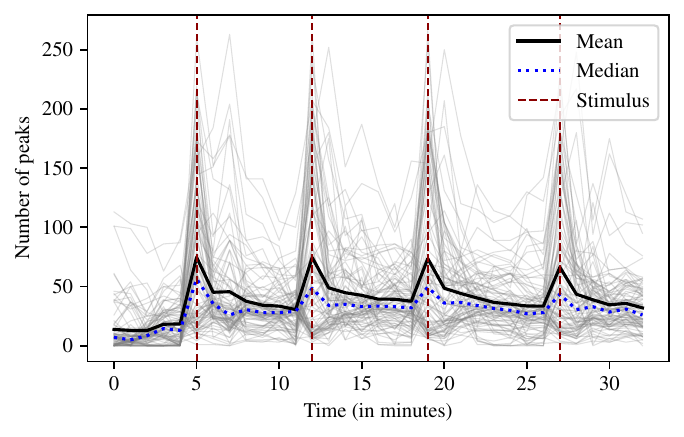}
\end{subfigure}
\begin{subfigure}{0.475\textwidth}
\centering
\caption{June 2023}
\includegraphics[width=\textwidth]{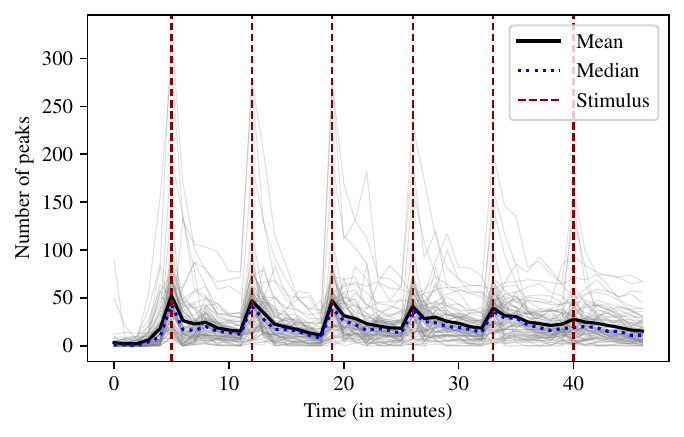}
\end{subfigure}
\begin{subfigure}{0.475\textwidth}
\centering
\caption{September 2023}
\label{fig:Sep1223}
\includegraphics[width=\textwidth]{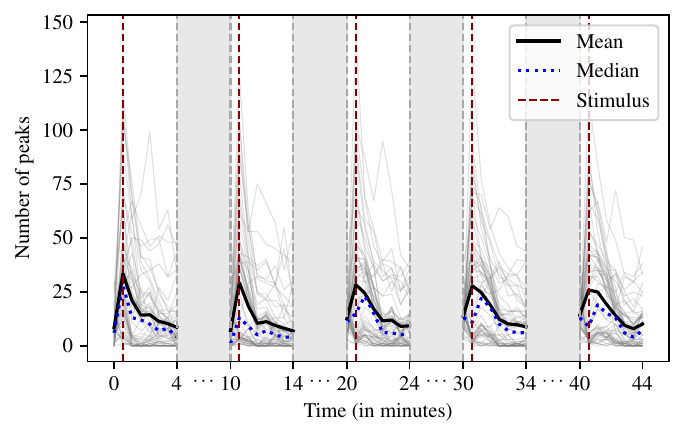}
\end{subfigure}
\begin{subfigure}{0.475\textwidth}
\centering
\caption{June 2025}
\label{fig:Jun0425}
\includegraphics[width=\textwidth]{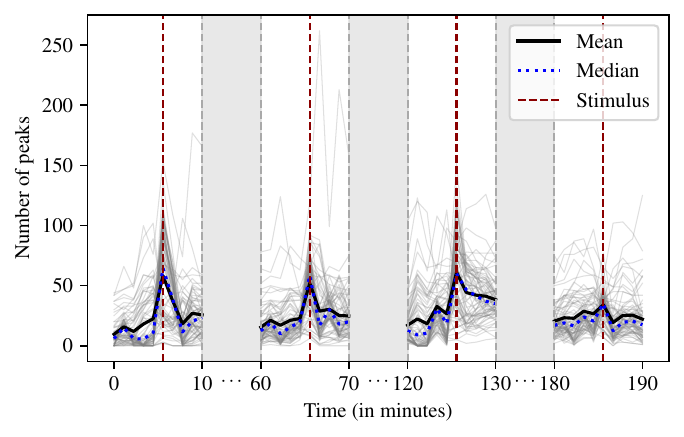}
\end{subfigure}
\caption{Number of recorded spikes per minute for each neuron across the four \textit{Aplysia} datasets, with corresponding mean and median values across all neurons for each recording. Intervals shaded grey correspond to censored periods with no observation.}
\label{fig:spikes}
\end{figure}

We consider four different experimental conditions, named after the date in which these were recorded: December 2022, June 2023, September 2023, and June 2025. Table~\ref{tab:data_summary} reports the duration of the recording, number of neurons involved, sampling rate, time of external stimuli, and length of breaks in recording for each experiment. 
Continuous recordings (December 2022 and June 2023 experiments) allow uninterrupted assessment of stimulus-driven responses, whereas recordings with short (10 minutes, September 2023 experiment) or long pauses (60 minutes, June 2025 experiment) enable the study of phenomena such as recovery and habituation to the effect of stimuli.
Additionally, Figure~\ref{fig:spikes} shows the  counts of recorded spikes per minute, along with their mean and median values, providing an overview of the firing intensity and its variability across neurons. 

\begin{table}[t]
\centering
\scalebox{0.975}{
\centering
\begin{tabular}{lccrcc}
\toprule
& Number of & Duration & Sampling & Stimuli (min) & Break length \\
Experiment & neurons & (min) & rate (Hz) & [excludes breaks] & (min) \\
\midrule
Dec 22 & 88 & 33 & 407.25 & 5, 12, 19, 27 & -- \\
Jun 23 & 77 & 47 & 407.25 & 5, 12, 19, 26, 33, 40 & -- \\
Sep 23 & 57 & 20 (5$\times$4) & 1629.00 & 0.5, 4.5, 8.5, 12.5, 16.5 & 6 \\
Jun 25 & 72 & 40 (4$\times$10) & 407.25 & 5, 15, 25, 35 & 50 \\
\bottomrule
\end{tabular}
}
\caption{Summary of the four \textit{Aplysia} neuronal spiking datasets and stimulation protocols.}
\label{tab:data_summary}
\end{table}

Importantly, Figure~\ref{fig:spikes} reveals several structural features that motivate the 
modelling framework proposed in this work. \emph{First}, firing rates exhibit clear increases following the application of external stimuli, indicating that event intensities are strongly stimulus-dependent and non-stationary. \emph{Second}, the magnitude of these increases varies across successive stimuli within the same experiment, suggesting adaptation effects such as attenuation over time. In some datasets, responses appear to weaken across repeated stimulation, consistent with gradual habituation. 
\emph{Third}, substantial heterogeneity is evident across neurons, both in baseline activity levels and in the strength and persistence of stimulus-driven responses. Some neurons display high baseline firing with small increases after a stimulus, whereas others exhibit low baseline activity but strong bursts following stimulation; in addition, the degree of attenuation or recovery to the baseline across repeated stimuli varies substantially between neurons. This heterogeneity suggests that a single set of excitation parameters is unlikely to adequately describe the population. In particular, this naturally motivates a clustering framework, whereby neurons are grouped according to shared hierarchical response patterns, enabling identification of latent subpopulations with distinct excitability 
structures.


\section{Hierarchical excitatory processes (HEP)}
\label{sec:hep}

Consider an increasing sequence of event times $t_1, t_2, \dots$ and let $N(t) = \sum_{i=1}^\infty \mathds{1}_{[0,t]}(t_i)$ denote the corresponding counting process, where $\mathds{1}_{\cdot}\{\cdot\}$ is the indicator function. In the \textit{Aplysia} example, the event times  represent the observed spikes 
for a single neuron. 
We assume that events occur according to an inhomogeneous Poisson point process with conditional intensity function $\lambda(t) : \mathbb{R}_{+} \to \mathbb{R}_{+}$, defined as 
$\lambda(t) = \lim_{\mathrm{d}t\to0} \mathbb{E}\left\{N(t+\mathrm{d}t) - N(t) \mid \mathcal{H}_t\right\} / \mathrm{d}t$, 
where $\mathcal{H}_t = \{t_i : t_i < t\}$ is the history of the process in $[0,t)$.
We assume that changepoints occur at a \textit{known} sequence of increasing time points $\tau_1,\tau_2,\dots$, corresponding to external 
stimuli applied to the 
process. 
Conditional on these changepoints, we model the conditional intensity $\lambda(t)$ at time $t\geq 0$ as
\begin{equation}
\lambda(t) = \lambda_0 + \sum_{j=1}^{\infty} \phi_{\boldsymbol{\theta}(\tau_j)}(t - \tau_j) \mathds{1}_{[0,t)}\{\tau_j\},
\label{eq:lambda_pp}
\end{equation}
where $\lambda_0 \in \mathbb{R}_{+}$ is the baseline intensity, and $\phi_{\boldsymbol{\theta}(s)}(\cdot) : \mathbb{R}_{+} \to \mathbb{R}_{+}$ is an excitation function parametrised by a $d$-dimensional, time-varying parameter vector $\boldsymbol{\theta}(s) = [\theta_1(s), \dots, \theta_d(s)]$ for $s\geq 0$. The excitation function quantifies how the occurrence of changepoints temporarily increases the event rate.
The parameters 
are themselves modelled via evolving functions of time $\theta_j:\mathbb{R}_{+} \to \mathbb{R},\ j=1,\dots,d$. For example, each parameter could be governed by an inhibitory 
mechanism:
\begin{equation}
\theta_j(s) = \max\left\{0, \theta_{0,j} - \sum_{i=1}^{\infty} \omega_j(s - \tau_i)  \mathds{1}_{[0,s)}\{\tau_i\} \right\},
\label{eq:theta_pp}
\end{equation}
where $\theta_{0,j}$ denotes the baseline parameter for $\theta_j$, and $\omega_j(\cdot)$ is a non-negative, monotonically decreasing inhibition function describing how repeated changepoints reduce subsequent values.
Alternatively, an excitatory Hawkes-type structure could be given, similarly to Equation~\eqref{eq:lambda_pp}: 
$\theta_j(t) = \theta_{0,j} + \sum_{i=1}^{\infty} \omega_j(t - \tau_i)  \mathds{1}_{[0,t)}\{\tau_i\}$. 
This two-level construction defines a \textit{hierarchical excitatory process (HEP)}, where the intensity $\lambda(t)$ depends on changepoints through an excitation structure, \textit{cf.}~Equation~\eqref{eq:lambda_pp}, and the parameters of this excitation are themselves dynamic and history-dependent via excitatory or inhibition latent functions, \textit{cf.}~Equation~\eqref{eq:theta_pp}.
The choice of an inhibition or excitatory structure for the parameter-level process depends on the application: for example, in the \textit{Aplysia} data described in Section~\ref{sec:data}, we expect repeated external stimuli to attenuate the neuronal response. Accordingly, an inhibitory structure as in Equation~\eqref{eq:theta_pp} is appropriate, allowing the excitation parameters to decrease after successive stimuli, capturing habituation effects in the firing dynamics.

In the literature for Hawkes processes, a common choice for both excitation and inhibition functions is the scaled exponential form:
\begin{equation}
\phi_{\boldsymbol{\theta}(s)}(u) = \alpha(s)  \exp\left\{-\beta(s)\, u\right\},
\label{eq:phi_exp}
\end{equation}
where $\alpha(s)$ is the jump magnitude and $\beta(s)$ controls the rate of decay following each changepoint. Similarly, for the inhibition of the parameters $\alpha(s)$ and $\beta(s)$, scaled exponential inhibition functions could be assumed, yielding the following parameter-specific functions:
\begin{gather}
\alpha(s) = \max\left\{0,  \theta_{0,\alpha} - \sum_{i=1}^{p} \eta_{\alpha}\exp\{-\xi_{\alpha}(s-\tau_i)\}  \mathds{1}_{[0,s)}\{\tau_i\}\right\}, 
\label{eq:alpha} \\
\beta(s) = \max\left\{0,  \theta_{0,\beta} - \sum_{i=1}^{p} \eta_{\beta}\exp\{-\xi_{\beta}(s-\tau_i)\}  \mathds{1}_{[0,s)}\{\tau_i\}\right\},
\label{eq:beta}
\end{gather}
where $\theta_{0,\cdot}, \eta_\cdot, \xi_\cdot\geq0$ 
respectively denote the baseline, jump, and decay parameters for the second-level Hawkes dynamics.

In the application to the \textit{Aplysia}'s central nervous system, a short delay can be expected between the stimulus and a jump in the intensity. Therefore, we consider an alternative excitation function with a delay $\delta$ before the jump $\alpha$, assuming that for the first $\delta$ units of time following a stimulus induced changepoint, the increase from $0$ to $\alpha$ is linear. This results in the following functional form: 
\begin{equation}
\phi_{\boldsymbol{\theta}(s)}(u) = \frac{\alpha(s) \, u}{\delta(s)}\mathds{1}_{[0,\delta(s))}\{u\} + \alpha(s) \exp\{-\beta(s)[u-\delta(s)]\}\left[1-\mathds{1}_{[0,\delta(s))}\{u\}\right]. \label{eq:phi_exp_delay}
\end{equation}
Similarly to $\alpha(s)$ and $\beta(s)$, the delay parameter $\delta(s)$ is adjusted after each changepoint via a scaled exponential inhibition function: 
\begin{equation}
    \delta(s) = \max\left\{0, \theta_{0,\delta} - \sum_{i=1}^{\infty} \eta_{\delta} \exp\{-\xi_{\delta} (s-\tau_i)\}\, \mathds{1}_{[0,s)}\{\tau_i\} \right\},
\end{equation}
where $\theta_{0,\delta}, \xi_\delta, \eta_\delta\geq0$. 

\begin{figure}[t]
\centering
\begin{subfigure}{0.475\textwidth}
\centering
\caption{Intensity function $\lambda(t)$}
\includegraphics[width=\textwidth]{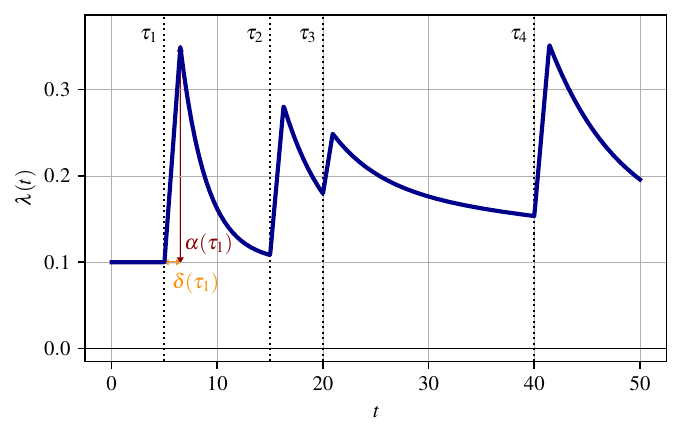}
\end{subfigure}
\begin{subfigure}{0.475\textwidth}
\centering
\caption{Jump parameter function $\alpha(t)$}
\includegraphics[width=\textwidth]{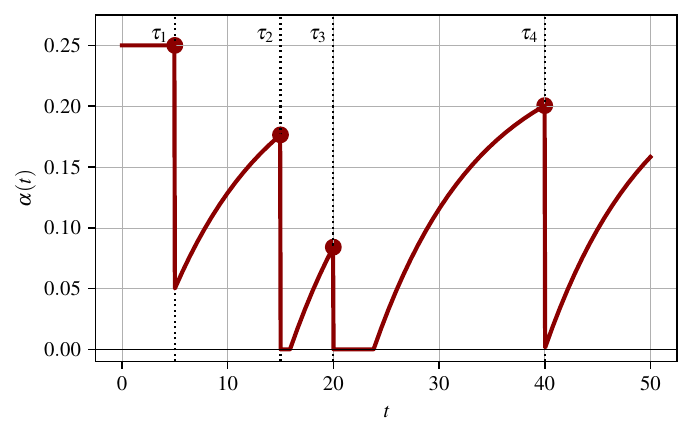}
\end{subfigure}
\begin{subfigure}{0.475\textwidth}
\centering
\caption{Decay parameter function $\beta(t)$}
\includegraphics[width=\textwidth]{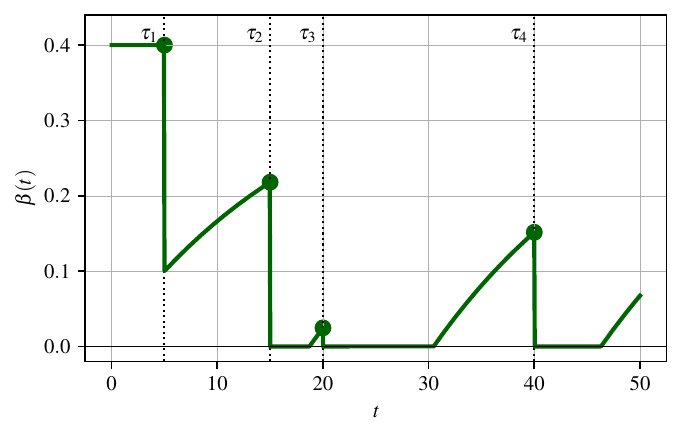}
\end{subfigure}
\begin{subfigure}{0.475\textwidth}
\centering
\caption{Delay parameter function $\delta(t)$}
\includegraphics[width=\textwidth]{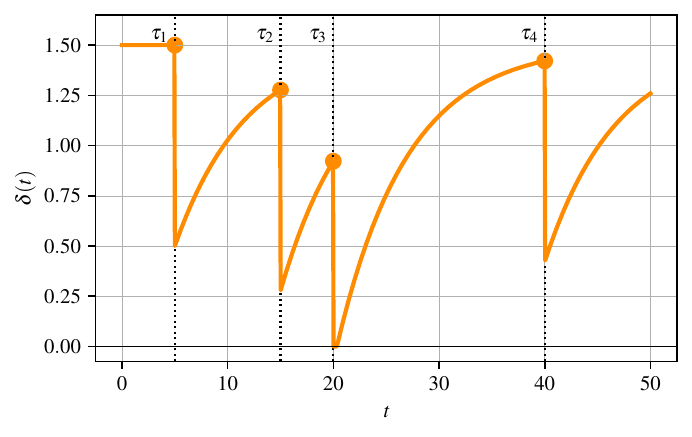}
\end{subfigure}
\caption{Example of intensity function $\lambda(t)$ in $[0,50]$ for an HEP with delayed scaled exponential excitation function, changepoints at $\tau_1=5,\ \tau_2=15,\ \tau_3=20,\ \tau_4=40$, and scaled exponential inhibition functions $\alpha(t)$, $\beta(t)$ and $\delta(t)$, with parameters $\lambda_0=0.1,\ \theta_{0,\alpha}=0.25,\ \eta_{\alpha}=0.2,\ \xi_{\alpha}=0.1,\ \theta_{0,\beta}=0.4,\ \eta_{\beta}=0.3,\ \xi_{\beta}=0.05,\ \theta_{0,\delta}=1.5,\ \eta_{\delta}=1.0,\ \xi_{\delta}=0.15$.}
\label{fig:toy}
\end{figure}

Figure~\ref{fig:toy} illustrates a toy example of the hierarchical excitatory process with the delayed scaled exponential excitation function in Equation~\eqref{eq:phi_exp_delay}, displaying the resulting intensity function $\lambda(\cdot)$, along with the underlying parameter functions $\alpha(\cdot)$, $\beta(\cdot)$, and $\delta(\cdot)$, respectively, for a sequence of four changepoints. The example demonstrates how the intensity evolves in response to successive changepoints and how the excitation, decay, and delay parameters are modulated over time by their corresponding inhibition functions, illustrating the class of intensity functions which can be modelled via HEPs.

\subsection{Parameter estimation in HEPs}

From observing event times $\boldsymbol{t} = (t_1, \dots, t_M)$ within an interval $[0, T]$, the log-likelihood under intensity $\lambda(t)$ is

\begin{equation}
\log L(\boldsymbol{\Theta}; \boldsymbol{t}) = \sum_{i=1}^{M} \log \lambda(t_i) - \int_0^T \lambda(u)\, \mathrm{d}u,
\label{eq:loglik}
\end{equation}
where $\boldsymbol{\Theta}$ collects all model parameters, including those of the inhibition functions for the time-varying parameters of the main HEP excitation function. For example, for the delayed scaled exponential excitation function 
in Equation~\eqref{eq:phi_exp_delay}, the set of all model parameters corresponds to $\boldsymbol{\Theta} = \{\lambda_0, \theta_{0,\alpha}, \eta_{\alpha}, \xi_{\alpha}, \theta_{0,\beta}, \eta_{\beta}, \xi_{\beta}, \theta_{0,\delta}, \eta_{\delta}, \xi_{\delta}\}$.
Principled estimates of the parameters $\boldsymbol{\Theta}$ can be obtained by maximising \eqref{eq:loglik} using, for example, gradient-based optimisation or quasi-Newton methods such as L-BFGS.

In practice, the integral in \eqref{eq:loglik} can be computed exactly for an HEP with conditional intensity function given by Equation~\eqref{eq:lambda_pp}: 
\begin{equation}
    \log L(\boldsymbol{\Theta}; \boldsymbol{t}) = \sum_{i=1}^{M} 
\log\left(
\lambda_0 + \sum_{i=1}^{M} \sum_{j=1}^{p} 
\phi_{\boldsymbol{\theta}(\tau_j)}(t_i - \tau_j)
\mathds{1}_{[0,t_i)}\{\tau_j\}
\right)
- \lambda_0 T 
- \sum_{j=1}^{p} I_j,
\label{eq:loglik_delay_full}
\end{equation}
where $p$ is the number of changepoint stimuli encountered in the observation window $[0,T]$ and $I_j=\int_{0}^{T - \tau_j} \phi_{\boldsymbol{\theta}(\tau_j)}(u)\, \mathrm{d}u$. The form of $I_j$ depends on the choice of the excitation function. For example, for the scaled exponential function in Equation~\eqref{eq:phi_exp}, we obtain $I_j = \alpha(\tau_j)\,\beta(\tau_j)^{-1} \left\{1 - e^{-\beta(\tau_j)\,(T - \tau_j)}\right\}$, whereas for the delayed scaled exponential function in Equation~\eqref{eq:phi_exp_delay}, the resulting integral is: 
\begin{equation}
I_j = 
\begin{cases}
\dfrac{\alpha(\tau_j)}{2\delta(\tau_j)} (T - \tau_j)^2, & \text{if } T - \tau_j \le \delta(\tau_j), \\[1em]
\dfrac{\alpha(\tau_j)\, \delta(\tau_j)}{2}
+ \dfrac{\alpha(\tau_j)}{\beta(\tau_j)}\left[1 - e^{-\beta(\tau_j) \{T - \tau_j - \delta(\tau_j)\}}\right],
& \text{if } T - \tau_j > \delta(\tau_j),
\end{cases}
\end{equation}

In some experiments, such as Figures~\ref{fig:Sep1223}~and~\ref{fig:Jun0425}, there may exist periods without recorded activity, due to the experimental design. In the general case with a set of observed intervals $[a_r, b_r],\ r=1,\dots,R$ within the window $[0,T]$,
the censored log-likelihood becomes
\begin{equation}
\log L_c(\boldsymbol{\Theta}; \boldsymbol{t}) = \sum_{i=1}^M \log \lambda(t_i) - \sum_{r=1}^R \int_{a_r}^{b_r} \lambda(u)\, \mathrm{d}u.
\end{equation}
This formulation allows the HEP model to handle experimental designs where certain periods are excluded from observation while still enabling likelihood evaluations on the observed portions of the timeline.


\subsection{Model-based clustering of HEPs}

As discussed in Section~\ref{sec:data}, in the \textit{Aplysia} example neurons often share similar response patterns, suggesting a similar underlying parametrisation when modelled as HEPs. Therefore, we now extend the HEP model to a model-based clustering framework, where neurons are assumed to belong to one of $K$ latent groups, each characterised by a distinct excitation–inhibition pattern. 
Specifically, each neuron $\ell\in\{1,\dots,n\}$ is associated with a latent variable $z_\ell \in \{1, \dots, K\}$, indicating its group membership. 
Conditional on $z_\ell=k$, the neuronal activity is modelled via a group-specific intensity function $\lambda^{(k)}(t)$, parametrised as
\begin{equation}
\lambda^{(k)}(t) = \lambda_{0}^{(k)} + \sum_{i=1}^{p} 
\phi_{\boldsymbol{\theta}^{(k)}(\tau_i)}(t - \tau_i)\,
\mathds{1}_{[0,t)}\{\tau_i\}, 
\qquad k = 1, \dots, K,
\label{eq:lambda_cluster}
\end{equation}
where, as before, $\lambda_{0}^{(k)}$ denotes the group-specific baseline rate and $\phi_{\boldsymbol{\theta}^{(k)}(\tau_i)}(\cdot)$ is the excitation function with group-specific parameters, which captures the impact of each stimulus at time $\tau_i$ on the subsequent firing intensity for a neuron in the $k$-th group. Note that the functional form of the excitation kernel is assumed to be the same across groups, but each group is associated with different model parameters. 
Intuitively, neurons assigned to the same group share similar temporal excitation patterns. For example, neurons in the same group may have comparable delays before a peak in the response, or similar rates of habituation across external stimuli. 
This clustering structure allows us to infer shared mechanisms of reaction to external stimuli, 
rather than modelling each one in isolation.

Inference for the clustering model in Equation~\eqref{eq:lambda_cluster} can be performed via an Expectation-Maximisation (EM) algorithm. For simplicity, we present the algorithm for the case where the entire window $[0,T]$ is observed, with the censored case being completely analogous. 
Let $\boldsymbol{t}_\ell = (t_{\ell,1}, \dots, t_{\ell,M_\ell})$ denote the observed spike times for neuron $\ell$, $\ell=1,\dots,n$. 
Each neuron belongs to one of the $K$ latent groups, indexed by $z_\ell \in \{1,\dots,K\}$, with prior mixing probabilities $\pi_k = \mathbb{P}(z_\ell = k)$, satisfying $\pi_k\geq0$ for all $k=1,\dots,K$ and $\sum_{k=1}^K \pi_k = 1$. 
The full log-likelihood, including group memberships, is
\begin{multline}
\log L_{\mathrm{full}}(\boldsymbol{\Theta}, \vec{z}; \vec t_1,\dots, \vec t_n) 
= \sum_{\ell=1}^n \sum_{k=1}^K 
\mathds{1}_{\{k\}}\{z_\ell\}
\Bigg[\log \pi_k
- \lambda_0^{(k)}T - \sum_{j=1}^p\int_0^{T-\tau_j} \phi_{\boldsymbol{\theta}^{(k)}(\tau_j)}(u)\, \mathrm{d}u \\
+
\sum_{i=1}^{M_\ell} \log\left(
\lambda_0 + \sum_{j=1}^{p} 
\phi_{\boldsymbol{\theta}^{(k)}(\tau_j)}(t_{\ell,i} - \tau_j)
\mathds{1}_{[0,t_{\ell,i})}\{\tau_j\}
\right)
\Bigg],
\label{eq:complete_ll_phi}
\end{multline}
where $\boldsymbol{\Theta} = \{\pi_k,\boldsymbol{\theta}^{(k)};\ k=1,\dots,K\}$ is the entire set of model parameters. The EM algorithm iteratively calculates estimates $\boldsymbol{\Theta}^{(r)}=\{\pi_k^{(r)}, \boldsymbol{\theta}^{(k,r)};\ k=1,\dots,K\}$ of $\boldsymbol{\Theta}$, alternating between an expectation step (E-step) and maximisation step (M-step).
At the $r$-th iteration, the E-step calculates the so-called responsibilities $\gamma_{\ell k}^{(r)}=\mathbb P(z_\ell = k \mid \vec t_\ell,\ \boldsymbol{\Theta}^{(r)})$, given current parameter estimates $\boldsymbol{\Theta}^{(r)}$. For the mixture of HEPs, the responsibilities take the form
\begin{equation}
\gamma_{\ell k}^{(r)} =  
\frac{\pi_k^{(r)} L(\boldsymbol{\theta}^{(k,r)}; \vec t_\ell)}
{\sum_{k^\prime=1}^K \pi_{k^\prime}^{(r)} L(\boldsymbol{\theta}^{(k^\prime,r)}; \vec t_\ell)},
\qquad \ell = 1,\dots,n,\ k = 1,\dots,K,
\label{eq:responsibility_phi}
\end{equation}
where $L(\cdot)$ is the likelihood function in Equation~\eqref{eq:loglik}.  
At the M-step, updated parameter estimates $\boldsymbol{\Theta}^{(r+1)}$ are obtained as: 
\begin{equation}
\boldsymbol{\Theta}^{(r+1)} = \argmax_{\boldsymbol{\Theta}}
\left[\sum_{\ell=1}^n \sum_{k=1}^K \gamma_{\ell k}^{(r)} 
\left\{
\log \pi_k
+ \log L(\boldsymbol{\theta}^{(k)}; \vec t_\ell)
\right\}\right].
\label{eq:Q_phi}
\end{equation}
For the mixing proportion, the closed-form solution to the update is: 
\begin{equation}
\pi_k^{(r+1)} = \frac{1}{n} \sum_{\ell=1}^n \gamma_{\ell k}^{(r)}.
\label{eq:pi_update_phi}
\end{equation}
For the remaining parameters, updates can be calculated numerically as follows: 
\begin{equation}
\boldsymbol{\theta}^{(k, r+1)} 
= \argmax_{\boldsymbol{\theta}} \left[\sum_{\ell=1}^n \gamma_{\ell k}^{(r)} 
\log L(\boldsymbol{\theta}; \vec{t}_\ell)\right],\quad k=1,\dots,K.
\label{eq:theta_update_phi}
\end{equation}

The EM iterations continue until the relative change in the observed-data log-likelihood $\log L_{\mathrm{obs}}^{(r)}$ falls below a tolerance or after a fixed number of iterations, where 
\begin{equation}
\log L_{\mathrm{obs}}^{(r)} 
= \sum_{\ell=1}^n 
\log \left\{
\sum_{k=1}^K 
\pi_k^{(r)} L(\boldsymbol{\theta}^{(k,r)}; \vec{t}_\ell)
\right\}. 
\end{equation}

Initial values for $\boldsymbol{\theta}^{(k)}$ can be obtained by first clustering neurons based on descriptive features 
or using specific algorithms for curve clustering based on a smoothed estimate of the intensity \citep[see, for example,][]{Heard06}. Conditional on the initial group allocations, the initial values $\boldsymbol{\Theta}^{(0)}$ are then calculated by fitting group-level HEPs separately.

The number of groups could be chosen via standard methods for model selection in mixture models, such as the integrated completed likelihood criterion
\citep[ICL;][]{Biernacki00}. The ICL can be approximated via the Bayesian information criterion (BIC) as follows: 
\begin{equation}
    \mathrm{ICL}(K) = \sum_{\ell=1}^n \log \left\{\sum_{k=1}^K \hat{\pi}_k L(\hat{\boldsymbol\theta}^{(k)}; \boldsymbol{t}_\ell)\right\} - \frac{\nu_K}{2} \log(n) + \sum_{\ell=1}^n \sum_{k=1}^K \hat\gamma_{\ell k} \log \hat\gamma_{\ell k},
    \label{eq:icl}
\end{equation}
where $\nu_K=\sum_{k=1}^{K}\mathrm{dim}(\boldsymbol{\theta}^{(k)})+K-1$ is the total number of model parameters, and $\hat{\gamma}_{\ell k}$ correspond to the responsibility for neuron $\ell$ in the $k$-th group, calculated with the final estimated values of the model parameters, \emph{cf.}~Equation~\eqref{eq:responsibility_phi}. 

\section{Simulations}
\label{sec:simulations}

In this section, we test the ability of the inferential procedure for HEPs to recover model parameters and underlying excitation and inhibition functions, and predict the intensity function for additional unseen stimuli on simulated data. Furthermore, we analyse the performance at recovering the latent group structure under different model specifications.

\subsection{Parameter estimation and prediction}
\label{sec:sim_1}

\begin{figure}[t]
\centering
\includegraphics[width=\textwidth]{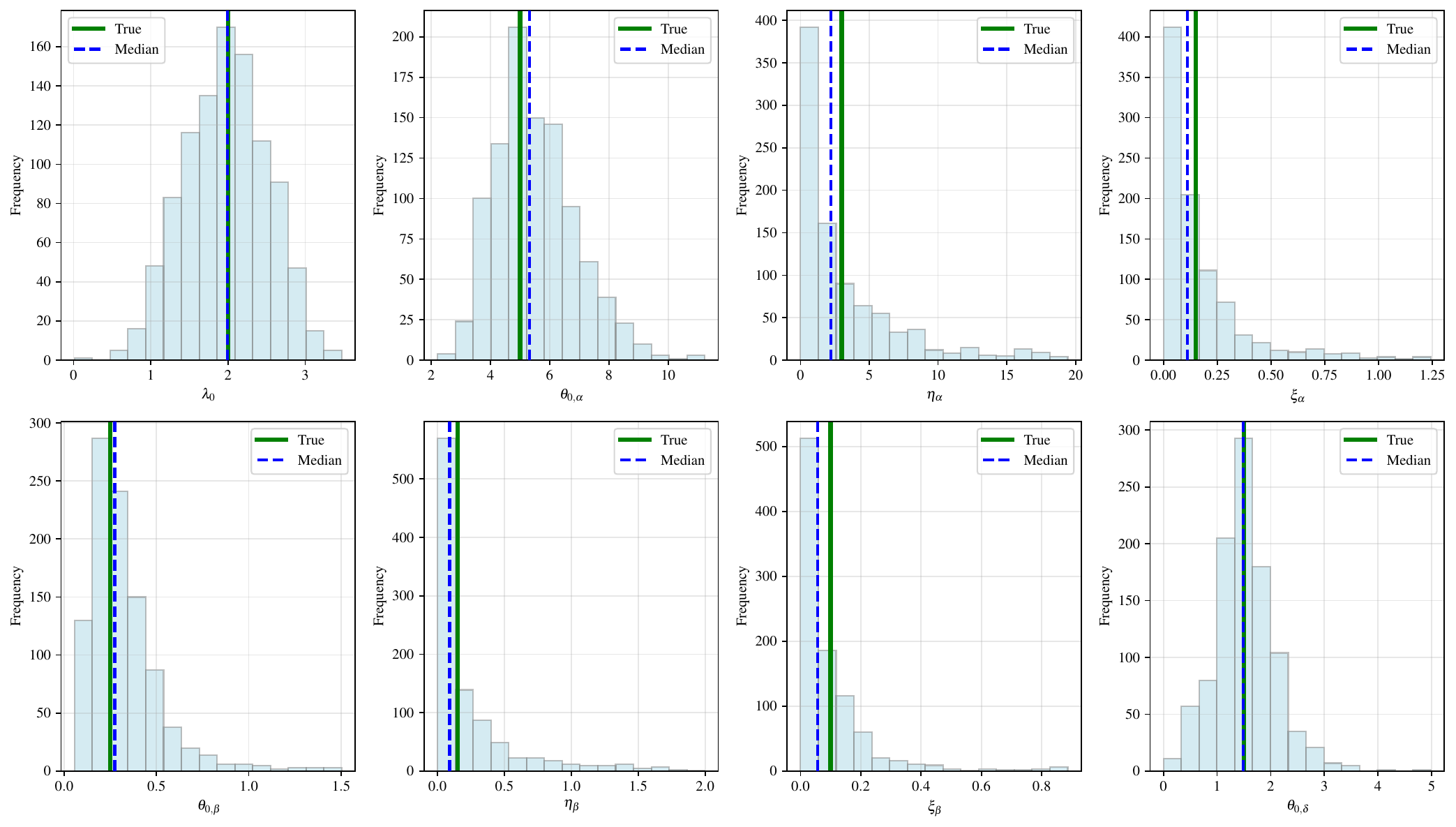}
\caption{Histogram of estimated parameters, corresponding medians across all $S=\numprint{1000}$ simulations, and true values for the synthetic data experiment described in Section~\ref{sec:sim_1}.}
\label{fig:sim1_histogram}
\end{figure}

\begin{figure}[t]
\centering
\begin{subfigure}{0.97\textwidth}
\centering
\caption{Intensity function $\lambda(t)$}
\includegraphics[width=0.975\textwidth]{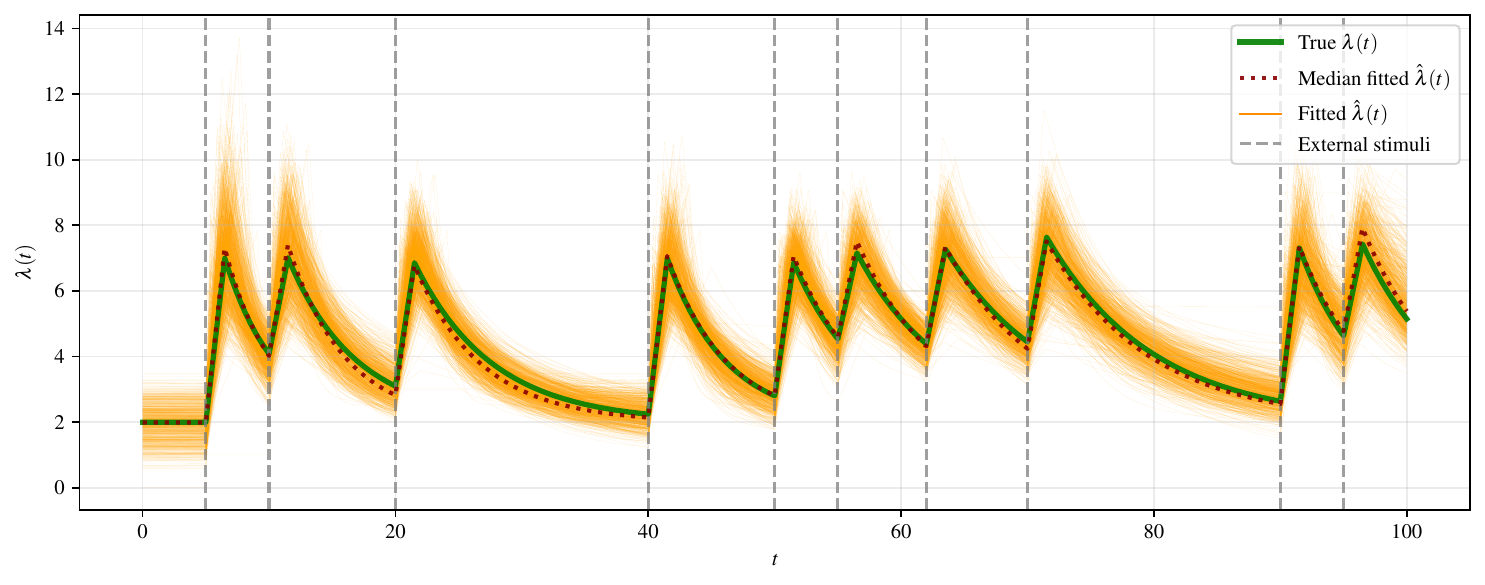}
\label{fig:sim1_intensities_lambda}
\end{subfigure}
\begin{subfigure}{0.475\textwidth}
\centering
\caption{Jump parameter function $\alpha(t)$}
\includegraphics[width=0.975\textwidth]{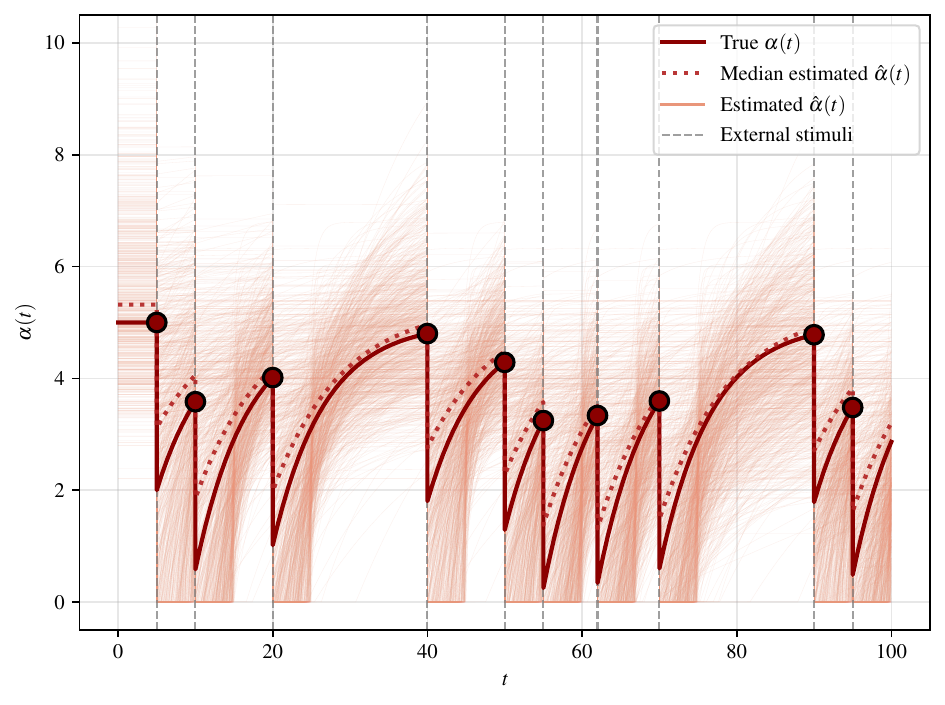}
\label{fig:sim1_intensities_alpha}
\end{subfigure}
\begin{subfigure}{0.475\textwidth}
\centering
\caption{Decay parameter function $\beta(t)$}
\includegraphics[width=0.975\textwidth]{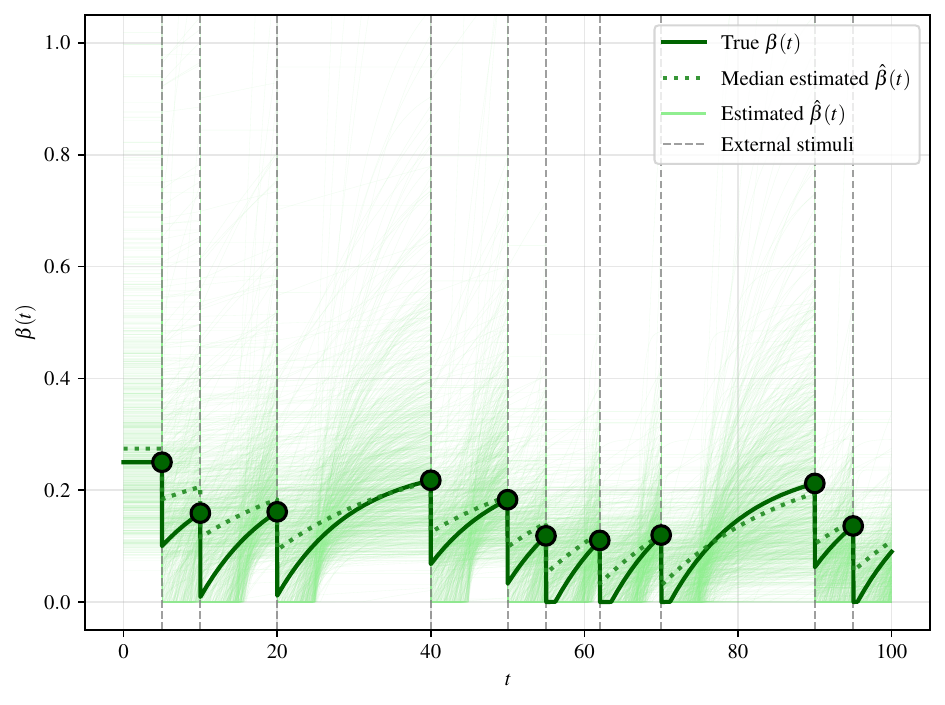}
\label{fig:sim1_intensities_beta}
\end{subfigure}
\caption{Estimated intensity functions $\lambda(\cdot)$ and hyperparameter functions $\alpha(\cdot)$ and $\beta(\cdot)$ across all $S=\numprint{1000}$ simulations, for the synthetic data experiment described in Section~\ref{sec:sim_1}.}
\label{fig:sim1_intensities}
\end{figure}

We simulate HEPs with intensity function taking the form detailed in Equation~\eqref{eq:lambda_pp}, with excitation function as in Equation~\eqref{eq:phi_exp_delay}. The jump and decay functions $\alpha(\cdot)$ and $\beta(\cdot)$ are given the same form as Equation~\eqref{eq:alpha} and~\eqref{eq:beta}, whereas the delay function $\delta(\cdot)$ is assumed constant across the entire observation period, corresponding to $\delta(\cdot)=\theta_{0,\delta}$ for some $\theta_{0,\delta}\in\mathbb{R}_+$. The parameter values in the simulation are set as follows: $\lambda_0=2.0$, $\theta_{0,\alpha}=5.0$, $\eta_{\alpha}=3.0$, $\xi_{\alpha}=0.15$, $\theta_{0,\beta}=0.25$, $\eta_{\beta}=0.15$, $\xi_{\beta}=0.1$, $\theta_{0,\delta}=1.5$. We set the total observation period to $T=100$, and we add $p=10$ external stimuli at locations $\boldsymbol\tau = (5, 10, 20, 40, 50, 55, 62, 70, 90, 95)$. We simulate data for $S=\numprint{1000}$ repetitions of the observation period, using the Lewis--Shedler thinning algorithm \citep{Lewis79}, and we perform inference on the model parameters for each simulation. To mitigate the effect of initialisation over the estimation procedure, 
we use three different initialisation schemes, and after convergence we pick the parameters corresponding to the highest log-likelihood. 
The results are reported in Figure~\ref{fig:sim1_histogram}, which plots the histograms of all model parameters across all simulations, and in Figure~\ref{fig:sim1_intensities}, which displays all estimated functions for the intensity, (Figure~\ref{fig:sim1_intensities_lambda}), jump (Figure~\ref{fig:sim1_intensities_alpha}) and decay parameters (Figure~\ref{fig:sim1_intensities_beta}) across the simulations. 

On average, all parameters appear to be recovered correctly, resulting in estimated intensity functions approximately tracking the true underlying intensity $\lambda(\cdot)$. For the hyperparameter functions $\alpha(\cdot)$ and $\beta(\cdot)$, performance appears to be particularly good at the locations of the external stimuli, possibly because the values $\alpha(\tau_i)$ and $\beta(\tau_i),\ i\in\{1,\dots,p\}$ appear directly in the equation for the intensity function $\lambda(\cdot)$. Considering that the hyperparameter functions only have access to $p=10$ noisy evaluations at the external stimuli, their average recovery in Figures~\ref{fig:sim1_intensities_alpha} and~\ref{fig:sim1_intensities_beta} is remarkably close to the underlying values, demonstrating a good performance of the inferential procedure even in the presence of a limited number of external stimuli.  

\begin{figure}[t]
\centering
\begin{subfigure}{0.97\textwidth}
    \caption{Intensity function $\lambda(t)$}
    \includegraphics[width=0.9506\textwidth]{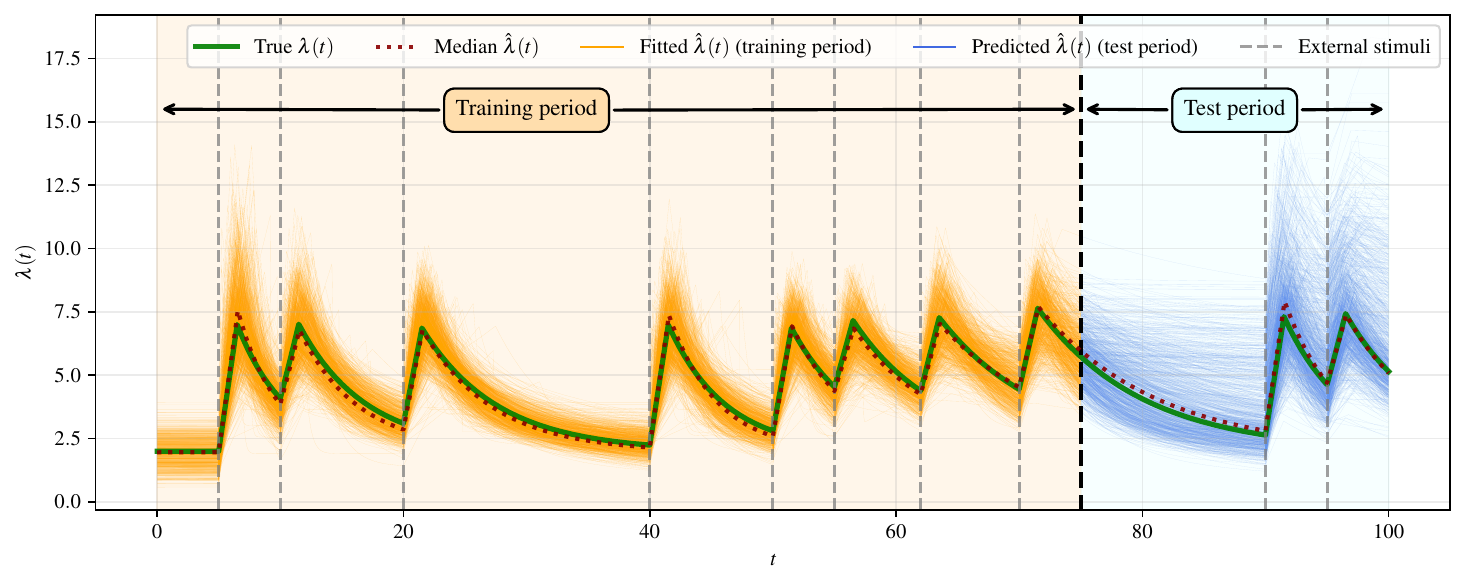}
\label{fig:sim1_intensities_prediction}
\end{subfigure}
\begin{subfigure}{0.97\textwidth}
    \caption{Standard deviation of the estimate of $\lambda(t)$}
    \includegraphics[width=0.9506\textwidth]{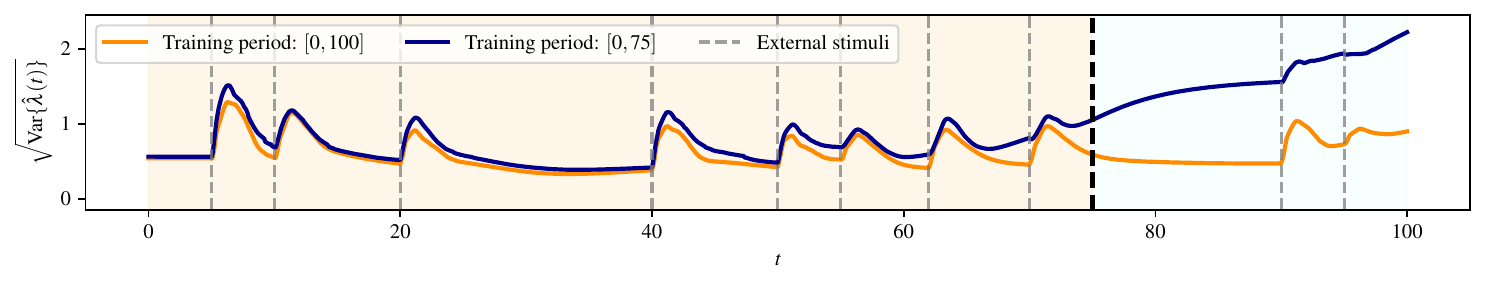}
\label{fig:sim1_intensities_variability}
\end{subfigure}
\caption{Estimated intensity functions $\lambda(\cdot)$ and standard deviation across all $S=\numprint{1000}$ simulations with parameters estimated on  the training period $[0,75]$, and predicted intensities in the test period $[75,100]$, for the synthetic data experiment described in Section~\ref{sec:sim_1}.}
\label{fig:sim1_intensities_pred}
\end{figure}

Next, we repeat the inference procedure focusing only on the time window $[0,75]$, which corresponds to six external stimuli, located at $\boldsymbol\tau = (5, 10, 20, 40, 50, 55, 62, 70)$. Then, we attempt to predict the intensity function in the window $(75,100]$, based on the model parameters obtained in the training period, introducing the additional external stimuli located at $\boldsymbol\tau^\ast = (90, 95)$. The resulting intensity estimates and predicted intensities are reported in Figure~\ref{fig:sim1_intensities_prediction}. As expected, Figure~\ref{fig:sim1_intensities_variability} shows that the variability of estimated intensities across simulations is slightly higher in the training period compared to Figure~\ref{fig:sim1_intensities}, as only $p=8$ external stimuli are used for estimation in this case, as opposed to $p=10$. Furthermore, predicted intensities in the test period show increased variability, but the underlying intensity is closely tracked using the median parameters estimated in the training period. 

\subsection{Clustering}
\label{sec:sim_2}

\begin{figure}[t]
\centering
\begin{subfigure}[t]{0.97\textwidth}
\centering
\caption{Group-specific intensity functions $\lambda^{(k)}(t),\ k\in\{1,2,3\}$}
\includegraphics[width=0.975\textwidth]{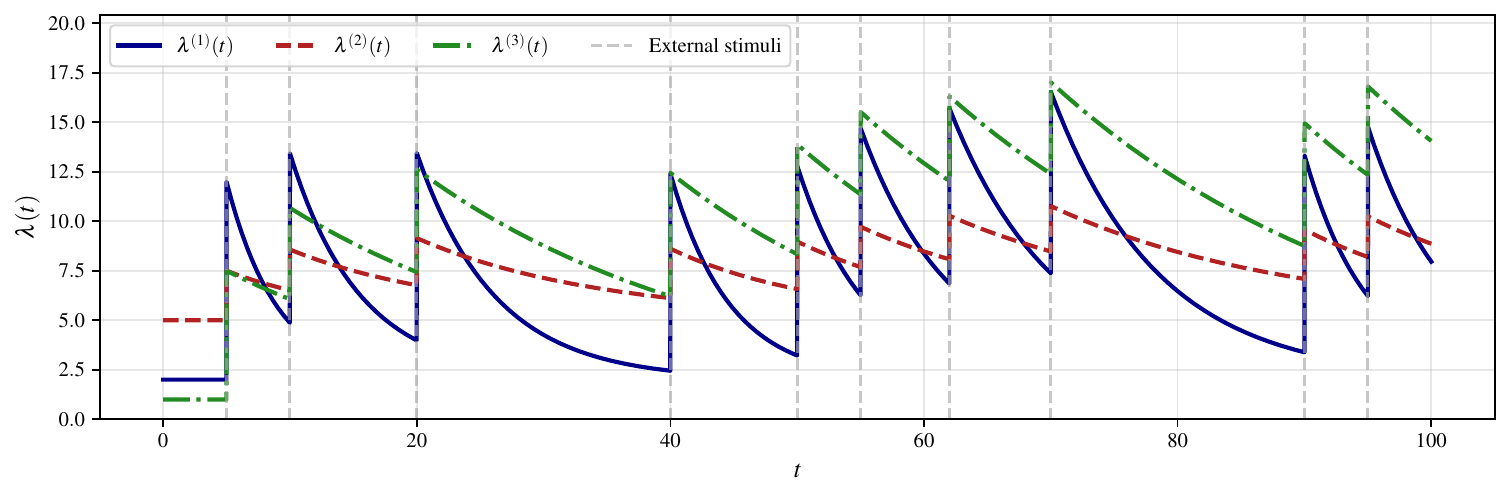}
\label{fig:sim2_intensities_lambda}
\end{subfigure}
\begin{subfigure}[t]{0.475\textwidth}
\centering
\caption{ARIs for different training periods $[0,T^\ast]$}
\includegraphics[width=0.975\textwidth]{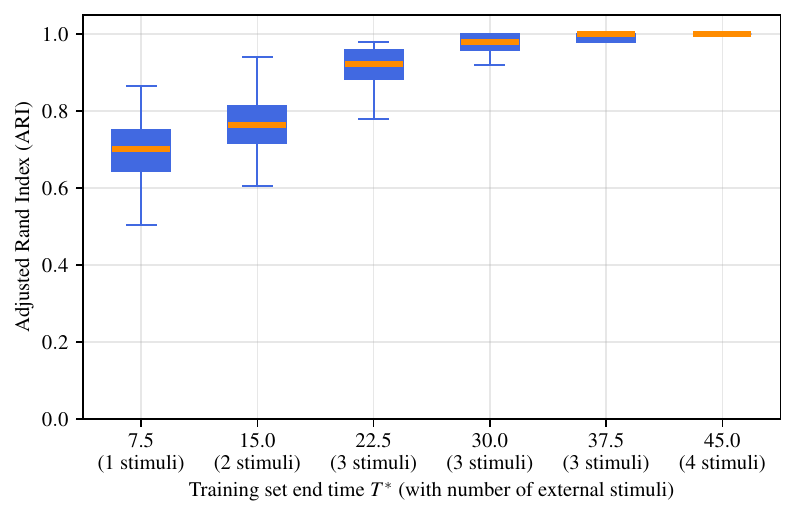}
\label{fig:sim2_ari}
\end{subfigure}
\begin{subfigure}[t]{0.475\textwidth}
\centering
\caption{$\hat{K}$ for different training periods $[0,T^\ast]$}
\includegraphics[width=0.975\textwidth]{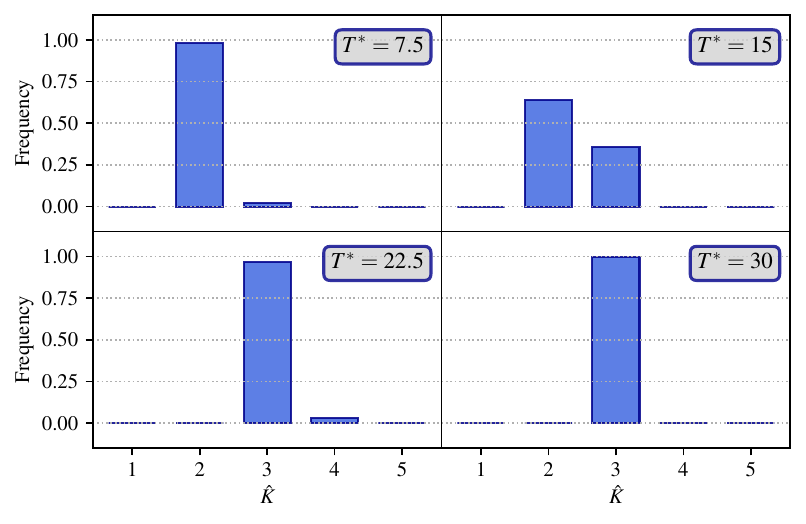}
\label{fig:sim2_number}
\end{subfigure}
\caption{Group-specific intensity functions for the synthetic data experiment described in Section~\ref{sec:sim_2}, and resulting ARIs and estimated number of groups for $S=\numprint{100}$ simulations.}
\label{fig:sim2_intensities}
\end{figure}

To test model-based clustering of event sequences, we simulate $n=150$ realisations of HEP processes divided equally into $K=3$ groups. 
The intensity function associated with each group takes the form in Equation~\eqref{eq:phi_exp}, with hyperparameter functions as in Equations~\eqref{eq:alpha} and~\eqref{eq:beta}, setting the same values for the observation period $T=100$ and the locations of external stimuli as Section~\ref{sec:sim_1} and with the following underlying parameters for each group: 

\begin{itemize}
    \item Group 1: $\lambda_0=2.0,\; \theta_{0,\alpha}=10.0,\; \eta_{\alpha}=5.0,\; \xi_{\alpha}=0.25,\; \theta_{0,\beta}=0.25,\; \eta_{\beta}=0.15,\; \xi_{\beta}=0.1$;
    
    \item Group 2: $\lambda_0=5.0,\; \theta_{0,\alpha}=2.5,\; \eta_{\alpha}=2.0,\; \xi_{\alpha}=0.3,\; \theta_{0,\beta}=0.1,\; \eta_{\beta}=0.1,\; \xi_{\beta}=0.15$;
    
    \item Group 3: $\lambda_0=1.0,\; \theta_{0,\alpha}=6.5,\; \eta_{\alpha}=4.0,\; \xi_{\alpha}=0.15,\; \theta_{0,\beta}=0.05,\; \eta_{\beta}=0.05,\; \xi_{\beta}=0.2$.
\end{itemize}
The resulting intensities are displayed in Figure~\ref{fig:sim2_intensities_lambda}. First, we fit a HEP mixture model with $K=3$ components, obtain estimates of the model parameters, and then estimate the group memberships as $\hat z_\ell=\argmax_{k\in\{1,\dots,K\}} \hat\gamma_{\ell k}$, where $\hat\gamma_{\ell k}$ are the responsibilities evaluated at the estimated values of the model parameters, \emph{cf.}~Equation~\eqref{eq:responsibility_phi}. We repeat the simulations $S=100$ times, and for each simulation we perform inference multiple times for $K=3$, restricting the observation intervals to time windows $[0,T^\ast]$ with $T^\ast\in\{7.5,15.0,22.5,30.0,37.5,45.0\}$, calculating estimated group memberships and the corresponding adjusted Rand index \citep{Hubert85} to evaluate the quality of the group structure recovery. Figure~\ref{fig:sim2_ari} displays the boxpots of ARI values across simulations, for different values of the end time $T^\ast$ in the training period. The plot also reports the total number of external stimuli observed in each interval $[0,T^\ast]$. Even when only one stimulus is observed, the inferential procedure distinguishes the group structure relatively well, reaching an almost perfect performance by $T^\ast=30$, shortly after the third external stimulus. This demonstrates a good performance of the proposed EM-algorithm at identifying the underlying groups correctly. 

Next, we fit HEP mixture models to the same simulated data, using a different number of components, ranging in $K\in\{1,\dots,6\}$, and we record the maximum value of the $\mathrm{ICL}(K)$ criterion, \textit{cf.}~Equation~\eqref{eq:icl}. 
Again, we repeat the inference procedure using different training intervals $[0,T^\ast]$, with $T^\ast\in\{7.5,15.0,22.5,30.0\}$.
Figure~\ref{fig:sim2_number} shows the barplots for the estimated number of groups $\hat{K}$ across $S=100$ simulations, obtained as $\hat{K} = \argmax_{K\in\{1,\dots,6\}}\mathrm{ICL}(K)$, for each value of $T^\ast$. The results in the plot demonstrate that, after an initial period (after only one or two stimuli) in which the number of groups is slightly underestimated, the correct number is correctly picked in the overwhelming majority of instances for subsequent external stimuli. This demonstrates a good performance of the chosen model selection criterion for identifying the underlying number of groups.

\section{Application to the \textit{Aplysia} neuronal spiking data}
\label{sec:applications}

\begin{figure}[t!]
\centering
\begin{subfigure}[t]{0.475\textwidth}
\centering
\includegraphics[width=0.975\textwidth]{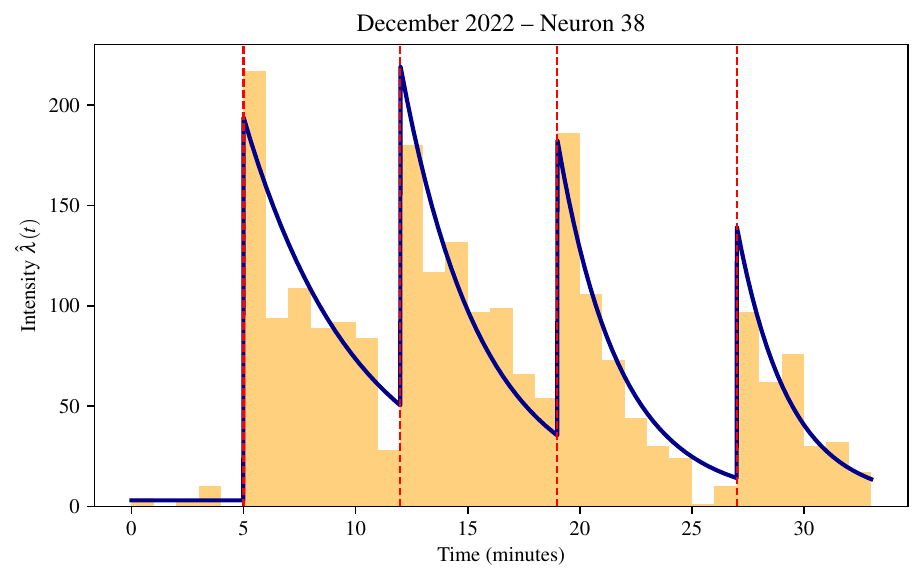}
\end{subfigure}
\begin{subfigure}[t]{0.475\textwidth}
\centering
\includegraphics[width=0.975\textwidth]{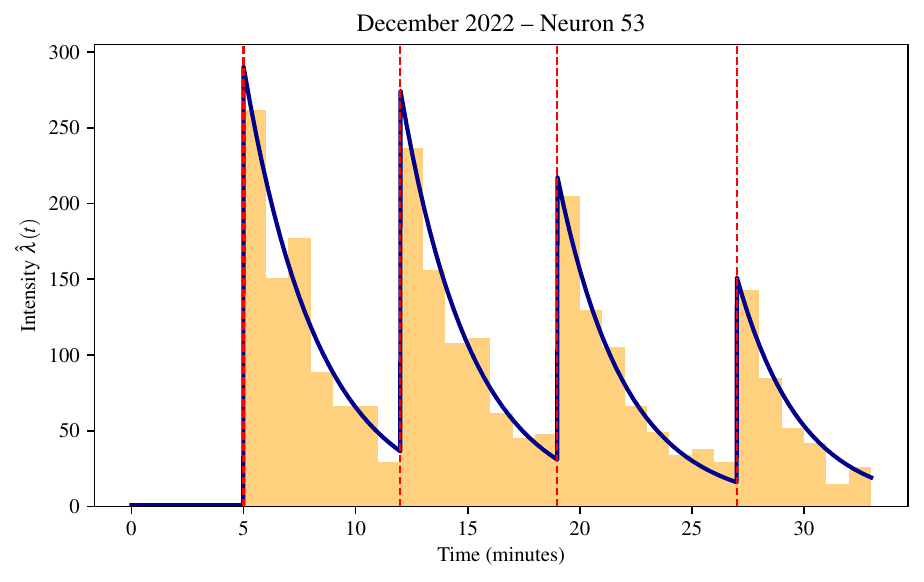}
\end{subfigure}
\begin{subfigure}[t]{0.475\textwidth}
\centering
\includegraphics[width=0.975\textwidth]{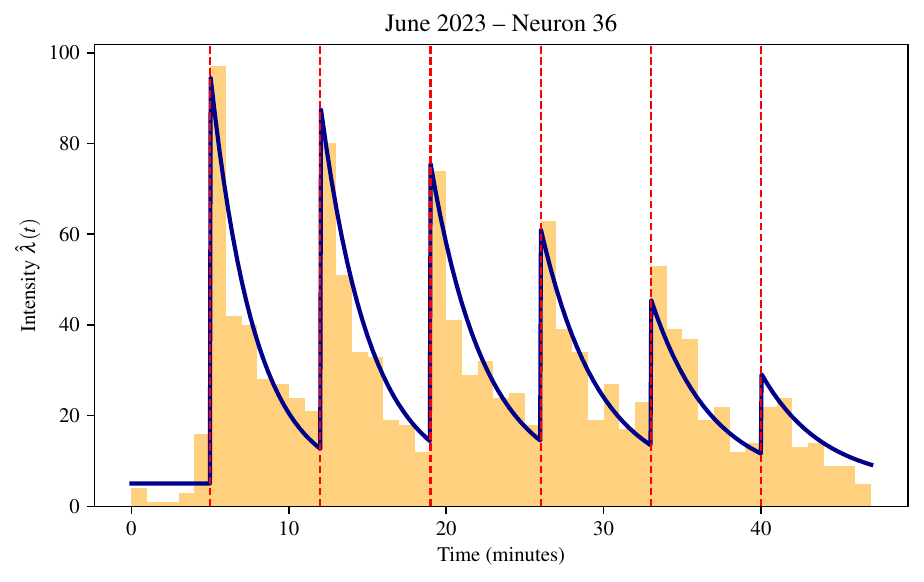}
\end{subfigure}
\begin{subfigure}[t]{0.475\textwidth}
\centering
\includegraphics[width=0.975\textwidth]{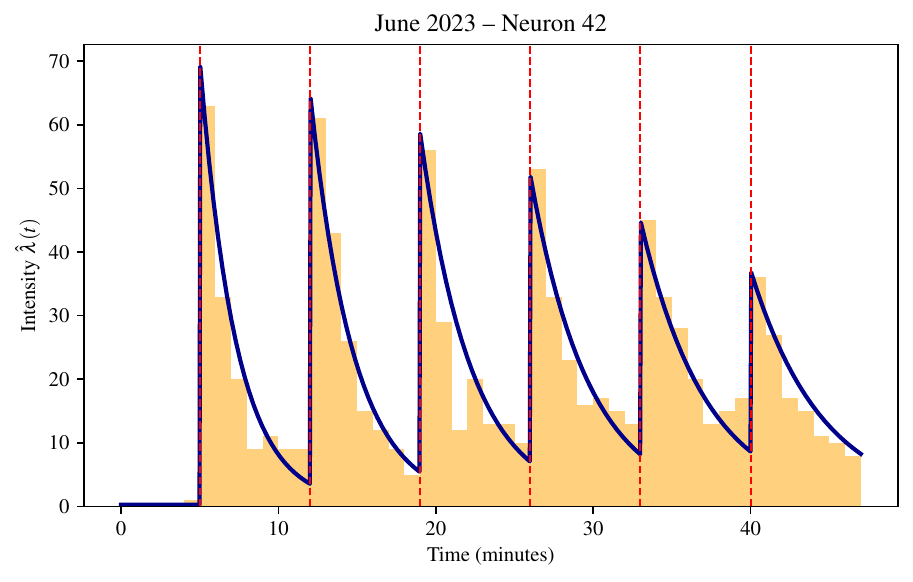}
\end{subfigure}
\begin{subfigure}[t]{0.475\textwidth}
\centering
\includegraphics[width=0.975\textwidth]{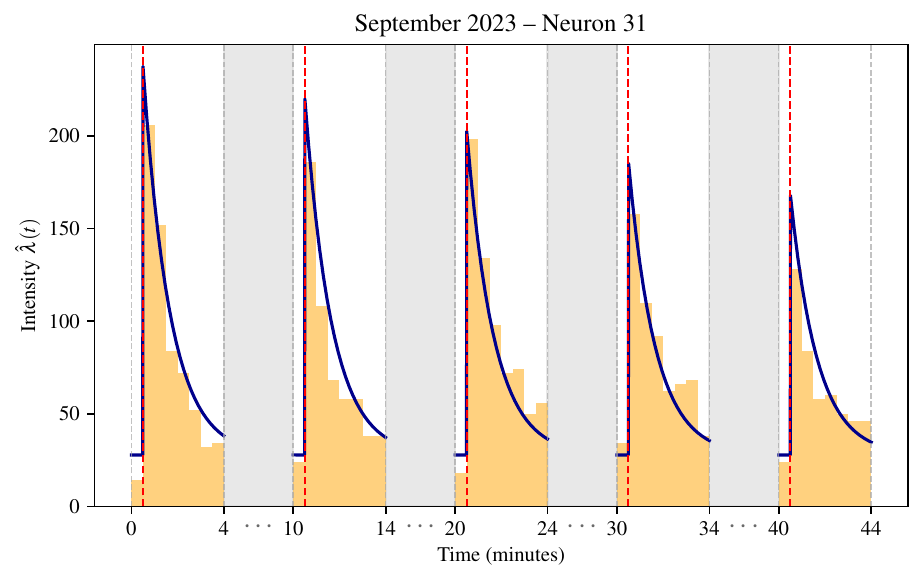}
\end{subfigure}
\begin{subfigure}[t]{0.475\textwidth}
\centering
\includegraphics[width=0.975\textwidth]{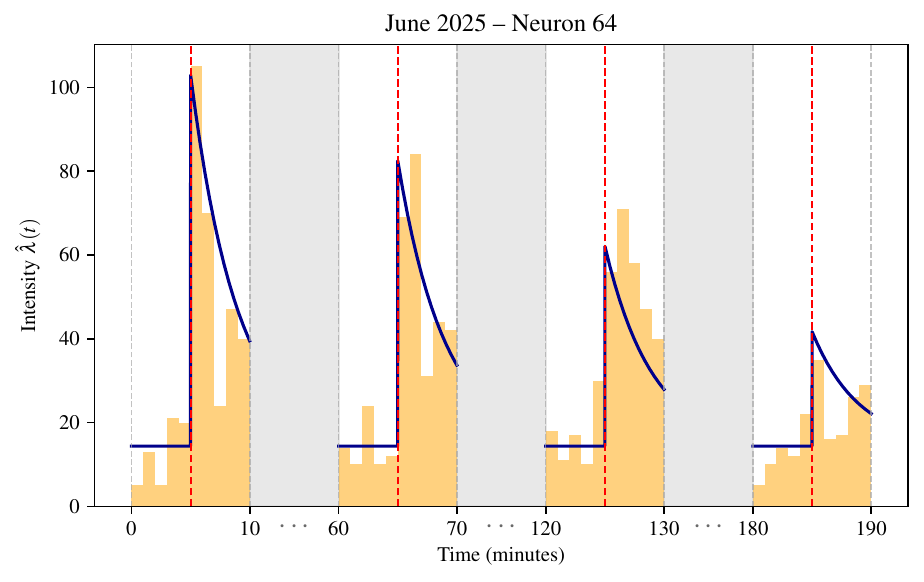}
\end{subfigure}
\caption{Fitted intensity functions and histograms of recorded spikes for six neurons in the \textit{Aplysia} data, under a HEP process with excitation function as in Equation~\eqref{eq:phi_exp}, with hyperparameter functions as in Equations~\eqref{eq:alpha} and~\eqref{eq:beta}. Note: $y$-axes are on different scales, and bin width for the histograms was set to one minute.}
\label{fig:individual_neurons}
\end{figure}

In this section, we apply the proposed methodology to the \textit{Aplysia} data described in Section~\ref{sec:data}. First, we fit HEP processes independently to each neuron across all experiments. We use three randomized initialisation schemes for the optimization procedure and retain the estimates related to the largest log-likelihood values. Figure~\ref{fig:individual_neurons} displays the estimated intensities and underlying event histograms for a selected subset of neurons. The intensity function associated with each model corresponds to Equation~\eqref{eq:phi_exp}, with hyperparameter functions as in Equations~\eqref{eq:alpha} and~\eqref{eq:beta}. The results show that the HEP model with varying jump sizes governed by an underlying Hawkes-like structure appears to be valid and an excellent fit for the observed firing times, conditional on the known external stimuli. Overall, the model appears to be particularly well suited to capturing spikes driven by known external stimuli, where the intensity of the jump gradually decreases based on the number of previous stimuli or their separation in time. 

\begin{figure}[t]
\centering
\begin{subfigure}[t]{0.325\textwidth}
\centering
\includegraphics[width=0.975\textwidth]{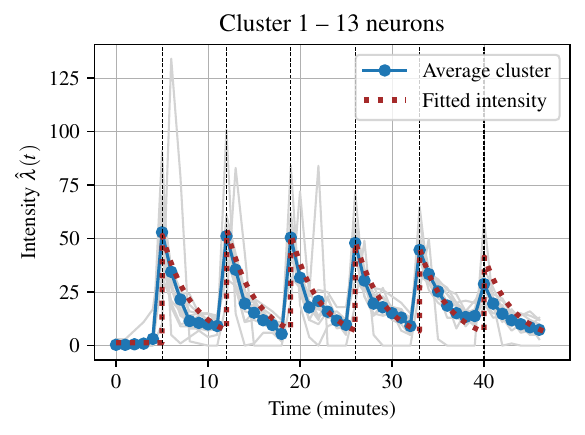}
\end{subfigure}
\begin{subfigure}[t]{0.325\textwidth}
\centering
\includegraphics[width=0.975\textwidth]{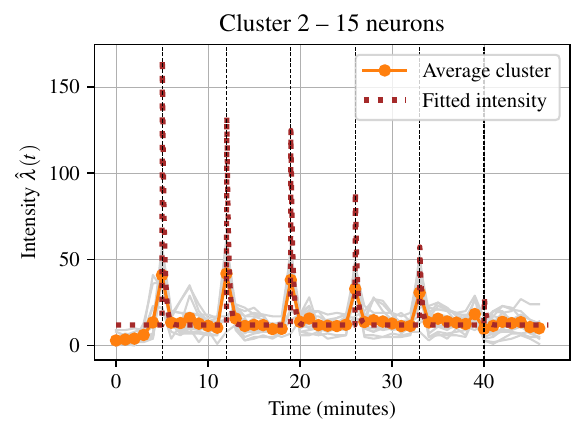}
\end{subfigure}
\begin{subfigure}[t]{0.325\textwidth}
\centering
\includegraphics[width=0.975\textwidth]{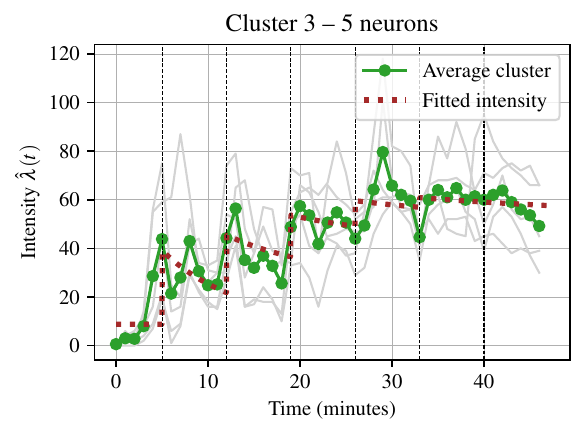}
\end{subfigure}
\centering
\begin{subfigure}[t]{0.325\textwidth}
\centering
\includegraphics[width=0.975\textwidth]{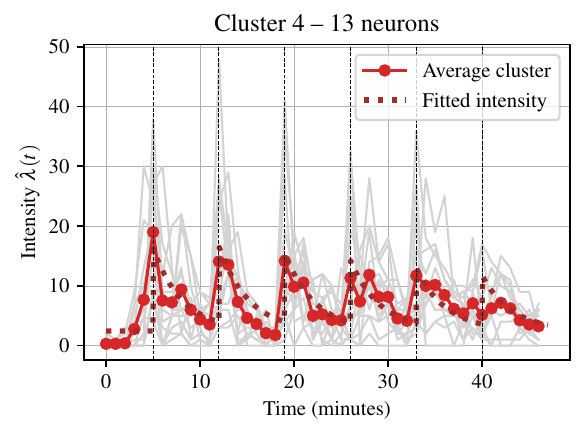}
\end{subfigure}
\begin{subfigure}[t]{0.325\textwidth}
\centering
\includegraphics[width=0.975\textwidth]{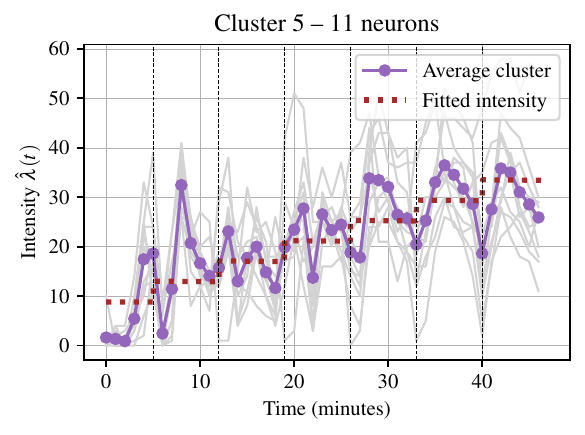}
\end{subfigure}
\begin{subfigure}[t]{0.325\textwidth}
\centering
\includegraphics[width=0.975\textwidth]{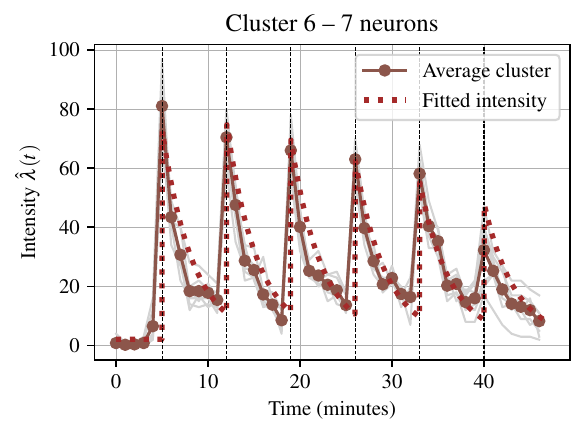}
\end{subfigure}
\centering
\begin{subfigure}[t]{0.325\textwidth}
\centering
\includegraphics[width=0.975\textwidth]{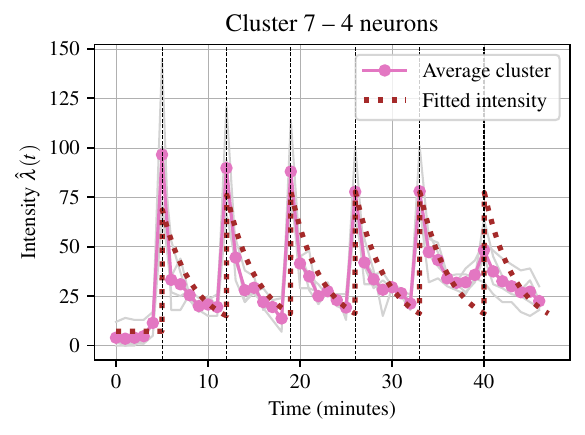}
\end{subfigure}
\begin{subfigure}[t]{0.325\textwidth}
\centering
\includegraphics[width=0.975\textwidth]{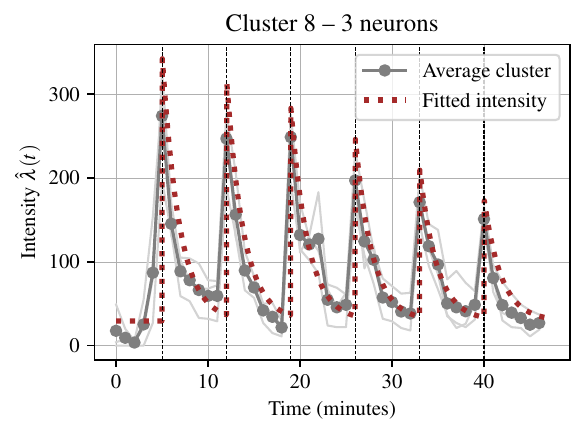}
\end{subfigure}
\begin{subfigure}[t]{0.325\textwidth}
\centering
\includegraphics[width=0.975\textwidth]{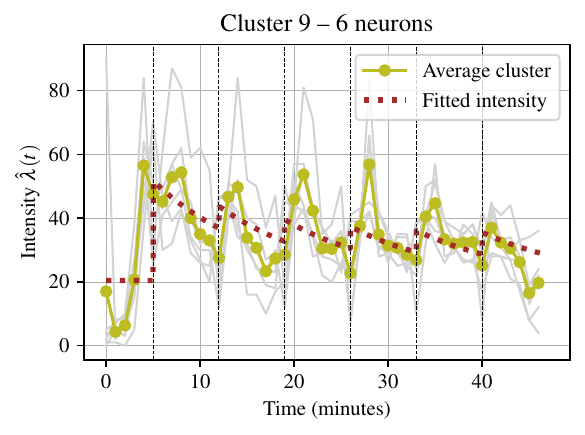}
\end{subfigure}
\caption{Average number of recorded spikes per minute and estimated intensity functions for the groups obtained via the HEP-based clustering with excitation function as in Equation~\eqref{eq:phi_exp}, with hyperparameter functions as in Equations~\eqref{eq:alpha} and~\eqref{eq:beta}, applied to the June 2023 \textit{Aplysia} data. Note: $y$-axes are on different scales.}
\label{fig:clustering}
\end{figure}

\begin{figure}[!t]
    \centering
    \includegraphics[width=0.995\linewidth]{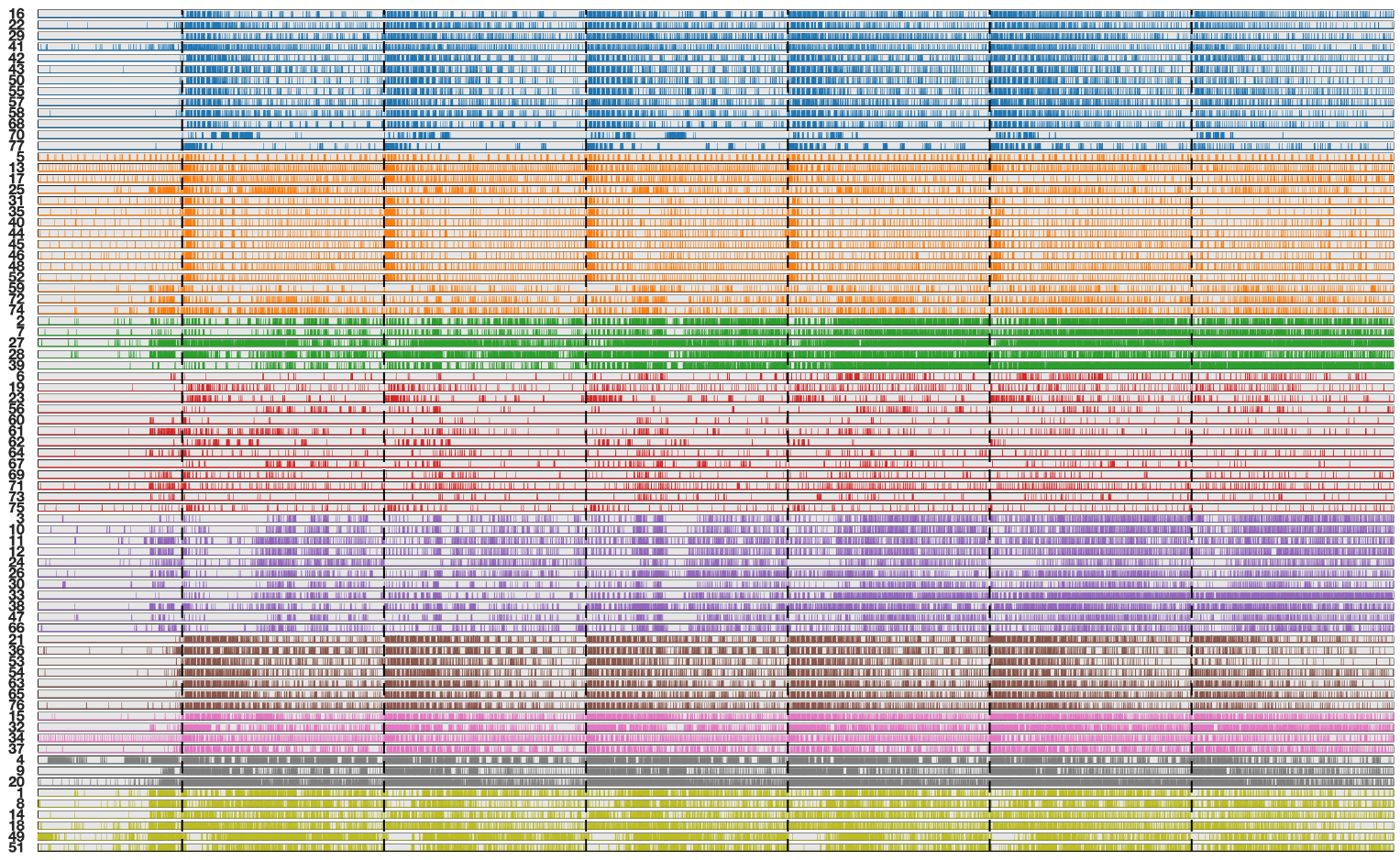}
    \caption{Electrophysiology traces displaying the spike-train activity of 77 unique neurons in the June 2023 \textit{Aplysia} data, organised according to HEP-based clustering. Color assignments for individual clusters correspond to those used in Figure~\ref{fig:clustering}. The broken vertical lines indicate when stimuli were applied (see Table~\ref{tab:data_summary}). }
    \label{fig:traces}
\end{figure}

Additionally, model-based clustering with HEPs can be used to discover groups of neurons that react similarly to external stimuli. 
Since the number of groups is unknown, we fit the HEP model for clustering with increasing values of $K$ (starting from $K=2$), until the value of $\mathrm{ICL}(K)$ stops increasing, selecting the resulting value as the optimal number of clusters $\hat{K}$. This procedure follows a technique similar to the methods used in popular packages for model-based clustering, such as \texttt{mclust} \citep[see, for example,][]{Scrucca16}. Figure~\ref{fig:clustering} displays the estimated cluster-specific intensities for the June 2023 experiment (where $\hat{K}=9)$, with the underlying event histogram for all neurons in that cluster. The corresponding electrophysiology traces are displayed in Figure~\ref{fig:traces}.

The results demonstrate that the HEP-based clustering is able to detect groups of neurons exhibiting similar response patterns to external stimuli. Note that the detected clusters are not limited to intensities closely resembling the examples in Figure~\ref{fig:individual_neurons}, but also to heterogeneous groups where the intensity does not decrease significantly after successive stimuli (such as groups 3 and 5), or does not spike after external interventions (such as group 9). The other inferred groups follow patterns similar to the structure identified in Section~\ref{sec:data} and Figure~\ref{fig:individual_neurons}, with jumps occurring at each external stimulus, with decreasing effects after successive interventions. The main differences between the groups relate to the underlying parameters associated with the HEP processes, which correspond to different effects of external stimuli on the neurons assigned to each group.   

\section{Conclusion and discussion}

In this work, we propose the Hierarchical Excitatory Process (HEP), a point process model for event-time data in the presence of multiple external stimuli at known locations in time. We propose hierarchical latent structures for the model parameters characterising the effects of external stimuli on the intensity function, and we discuss likelihood-based approaches to inference and model-based clustering. We apply HEPs to spike train recordings obtained from the sea slug \textit{Aplysia}'s pedal ganglion under repeated external stimulation, demonstrating good performance in estimating the effect of interventions on each neuron. 

The proposed method has some limitations: first, we focus on inhomogeneous Poisson processes for mathematical tractability, which may not be sufficiently rich to capture the complexity of some real-world observed processes. Possible extensions of HEPs could incorporate self-exciting and mutually-exciting multivariate processes, under a putative network structure \citep[see, for example][]{Linderman14, SannaPassino23}.  
Second, the intensities associated with each neuron are modelled independently. If additional information is available on the relation between neurons and their locations within the brain, it would be possible to utilize derived covariates to estimate network effects in the intensity functions \citep[see, for example,][]{SannaPassino24}. 
However, we demonstrate in this work how HEPs provide a simple and flexible framework which can provide practitioners with useful tools for understanding the effect of external interventions on point process data, and help predict the effect of future exogenous stimuli. Additionally, when covariates associated with the external stimuli are available, the excitatory and inhibitory functions in HEPs could also be easily adjusted. For example, in the \textit{Aplysia} example, HEP could be adapted to have the capacity to negotiate stimuli of different intensities applied within the same experiment.

\section*{Code availability}

Code to reproduce all simulation and experiments in this work is available in the Github repository \href{https://github.com/fraspass/hep}{\texttt{fraspass/hep}}.
\section*{Acknowledgements}

Francesco Sanna Passino acknowledges funding from the Engineering and Physical Sciences Research Council (EPSRC), grant number EP/Y002113/1. Nicholas A. Heard acknowledges funding from the EPSRC, grant number EP/X002195/1. William N. Frost is supported by the grant R01NS121220 from the National Institutes of Health (NIH).
Vince P. Lyzinski is supported by the Air Force Office of Scientific Research (AFOSR) Complex Networks award number FA9550-25-1-0128, and The Johns Hopkins University HLT COE.

\bibliographystyle{abbrvnat_mod}
\bibliography{reference}

\end{document}